\documentclass[twocolumn,showpacs,aps,preprintnumbers,amsmath,amssymb]{revtex4}

\usepackage{graphicx,subfigure}
\usepackage{dcolumn}
\usepackage{bm}
\usepackage{axodraw4j}

\makeatletter
\newcommand{\fmslash}[2][0mu]{%
  \mathchoice
    {\fmsl@sh\displaystyle{#1}{#2}}%
    {\fmsl@sh\textstyle{#1}{#2}}%
    {\fmsl@sh\scriptstyle{#1}{#2}}%
    {\fmsl@sh\scriptscriptstyle{#1}{#2}}}
\newcommand{\fmsl@sh}[3]{%
  \m@th\ooalign{$\hfil#1\mkern#2/\hfil$\crcr$#1#3$}}
\makeatother

\newcommand{\met}{{\fmslash E_T}}

\begin{document}


\title{A general method for determining the masses \\ 
of semi-invisibly decaying particles at hadron colliders}

\author{Konstantin T.~Matchev and Myeonghun Park}
\affiliation{Physics Department, University of Florida, Gainesville, FL 32611, USA}

\date{27 December, 2010}

\begin{abstract}
We present a general solution to the long standing problem of determining the masses
of pair-produced, semi-invisibly decaying particles at hadron colliders. 
We define two new transverse kinematic variables, $M_{CT_\perp}$ and $M_{CT_\parallel}$,
which are suitable one-dimensional projections of the contransverse mass
$M_{CT}$. We derive analytical formulas for the 
boundaries of the kinematically allowed regions in the 
$(M_{CT_\perp},M_{CT_\parallel})$ and $(M_{CT_\perp},M_{CT})$ parameter planes,
and introduce suitable variables $D_{CT_\parallel}$ and $D_{CT}$ 
to measure the distance to those boundaries on an event per 
event basis. We show that the masses can be reliably extracted from the 
endpoint measurements of 
$M_{CT_\perp}^{max}$ and $D_{CT}^{min}$ (or $D_{CT_\parallel}^{min}$).
We illustrate our method with dilepton $t\bar{t}$ events at the LHC.
\end{abstract}

\maketitle

The ongoing run of the Large Hadron Collider (LHC) at CERN will finally
provide the first glimpse of physics at the TeV scale.
In large part, the excitement surrounding the LHC is fueled by the
anticipation of the unknown: no one knows for sure where or how
the first signal of new physics beyond the standard model (BSM)
will show up. Yet, complementary and independent
arguments from particle physics {\em and} astrophysics
suggest that the best place to look for new physics
is a channel with missing transverse energy $\met$, 
caused by unseen new particles contributing to the 
dark matter of the Universe.

Unfortunately, the study of missing energy signatures poses 
a tremendous challenge at hadron colliders like the LHC.
The first fundamental difficulty is related to the 
very nature of hadron colliders, 
where in each event the partonic center-of-mass 
energy $\sqrt{\hat{s}}$ and longitudinal momentum 
$p_z$ of the initial state are unknown.
To make matters worse, the lifetime of the dark matter particle
is typically protected by a new parity symmetry,
which guarantees that in every event the missing
particles come in pairs, thus proliferating the number 
of unknown parameters describing the final state event 
kinematics.  

\begin{figure}[b!]
\includegraphics[width=6.5cm]{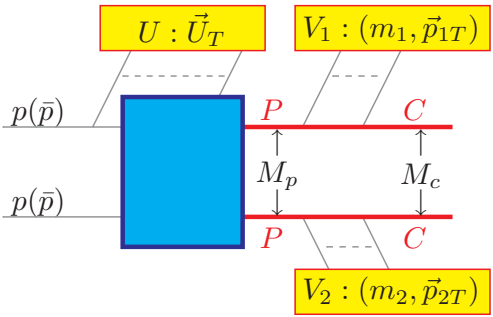}
\caption{\label{fig:event} 
The generic event topology under consideration.
All particles visible in the detector are clustered into three groups:
upstream objects $U$ with total transverse momentum $\vec{U}_T$,
and two composite visible particles $V_i$ ($i=1,2$), 
each with invariant mass $m_i$ 
and total transverse momentum $\vec{p}_{iT}$.
}
\end{figure}

The generic topology of a ``new physics'' $\met$ event
is sketched in Fig.~\ref{fig:event}.
Consider the {\em inclusive} production of an identical pair of 
new ``parent'' particles $P$. Each parent $P$ decays 
semi-invisibly to a set $V_i$ ($i=1,2$)
of standard model (SM) particles, which are visible in the detector, 
and a dark matter particle $C$ (from now on referred to as the ``child'') 
which escapes detection. In general, the parent pair is accompanied by
a number of additional ``upstream'' objects $U$ (typically jets)
with total transverse momentum $\vec{U}_T$. They
may originate from various sources such as initial state radiation 
or decays of even heavier particles.
We shall not be interested in the exact details of the 
physics responsible for $U$, adopting a fully inclusive approach 
to the production of the parents $P$. Given this general setup, 
the goal is to determine {\em independently} the mass $M_p$ of the parent
and the mass $M_c$ of the child in terms of $U$, $V_1$ and $V_2$.

In the past, several approaches to this problem have been proposed, 
but each has its own limitations.
For example, the classic method of invariant 
mass endpoints \cite{imass,Matchev:2009iw} only applies when the visible SM 
particles in $V_i$ arise from a sufficiently long decay chain. 
Attempts at direct reconstruction \cite{exactreco} of the 
children momenta are again limited to long decay chains only. 
In this letter, we shall consider the extreme, most challenging example
where each visible set $V_i$ consists of a {\em single}
SM particle of fixed mass $m_i$. A perfect testing ground 
for this scenario is provided by dilepton $t\bar{t}$ events
(already observed at the LHC \cite{Khachatryan:2010ez})
and we shall use that example in our numerical illustrations below.
The role of the parent $P$ (child $C$) will be played by
the SM $W$-boson (SM neutrino), each $V_i$ is a SM lepton
($e$ or $\mu$),
while $U$ is composed of the two $b$-jets from the top 
quark decays, plus any additional QCD jets from initial 
state radiation (ISR).

For such extremely short decay chains, the only viable alternative 
at the moment is provided by the methods based 
on the $M_{T2}$ variable \cite{approxreco}.
There, at least in principle, 
the individual masses $M_p$ and $M_c$
can be determined by observing a ``kink'' 
feature in the $M_{T2}$ endpoint as a function 
of a hypothesized trial mass $M_c$ for $C$ \cite{kink}, 
or by exploring the $U_T$ dependence of the $M_{T2}$ 
endpoint \cite{Matchev:2009fh}. 
Compared to those $M_{T2}$ approaches, our method here
has two advantages. First, it is simpler -- it
uses only the observed objects $U$, $V_1$ and $V_2$ 
in the event and makes no reference to the missing 
particle kinematics (or mass).
Second, it is more precise, since it utilizes  
the whole kinematic boundary of the relevant 
{\em two-dimensional} distribution and not just
the kinematic endpoint of its {\em one-dimensional} projection.
We proceed in three easy steps.

\begin{figure}[t!]
\includegraphics[width=6.5cm]{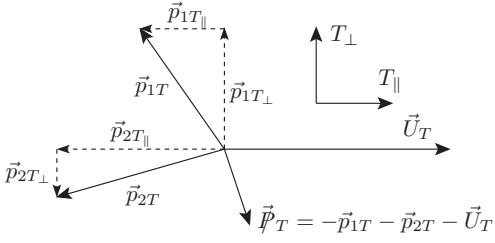}
\caption{\label{fig:trans} 
Decomposition of the observed transverse 
momentum vectors from Fig.~\ref{fig:event} in the transverse plane.
}
\end{figure}

{\em Step I. Orthogonal decomposition of the 
observed transverse momenta with respect to the 
$\vec{U}_T$ direction.} 
The Tevatron and LHC collaborations currently use 
fixed axes coordinate systems to describe their data.
Instead, we propose to rotate the coordinate system 
from one event to another, so that the transverse 
axes are always aligned with the direction $T_\parallel$
selected by the measured upstream transverse momentum vector
$\vec{U}_T$ and the direction $T_\perp$ orthogonal 
to it (see Fig.~\ref{fig:trans}). 
The visible transverse momentum vectors from Fig.~\ref{fig:event}
are then decomposed as
\begin{eqnarray}
\vec p_{iT_\parallel} &\equiv& \frac{1}{U_T^2}\left(\vec{p}_{iT}\cdot\vec{U}_T\right)\vec{U}_T ,
\label{pitpar} \\
\vec p_{iT_\perp} &\equiv& \vec{p}_{iT}-\vec p_{iT_\parallel}
= \frac{1}{U_T^2} \vec{U}_T \times \left(\vec{p}_{iT}\times \vec{U}_T\right).
\label{pitperp}
\end{eqnarray}

{\em Step II. Constructing the transverse and longitudinal 
contransverse masses $M_{CT_\perp}$ and $M_{CT_\parallel}$.} 
Our starting point is the original
contransverse mass variable \cite{Tovey:2008ui}
\begin{equation}
M_{CT}=\sqrt{m_1^2+m_2^2+2\left(e_{1T} e_{2T}
                +\vec p_{1T}\cdot \vec p_{2T}\right)},
\label{eq:mct}				
\end{equation}
where $e_{iT}$ is the ``transverse energy'' of $V_i$
\begin{equation}
e_{iT} = \sqrt{m_i^2+|\vec p_{iT}|^2}.
\end{equation}
For events with $U_T=0$, $M_{CT}$ has an upper endpoint which is 
insensitive to the unknown $\sqrt{\hat s}$,
providing one relation among $M_p$ and $M_c$ \cite{Tovey:2008ui,Polesello:2009rn}
\begin{equation}
M_{CT}^{max}(U_T=0) = \sqrt{m_1^2+m_2^2+2m_1 m_2
\cosh\left(\zeta_1+\zeta_2\right)},
\label{MCTmaxPT0}
\end{equation}
where 
\begin{eqnarray}
\sinh{\zeta_i} &\equiv& \frac{\lambda^{\frac{1}{2}}(M_p^2,M_c^2,m_i^2)}{2 M_p m_i}\, ,
\label{zetadef} \\ 
\lambda(x,y,z) &\equiv& x^2+y^2+z^2 - 2xy - 2xz - 2yz\, .
\end{eqnarray}

Unfortunately, the $U_T=0$ limit is not particularly interesting
at hadron colliders (especially for inclusive studies),
since a significant amount of upstream $U_T$ is
typically generated by ISR (and other) jets.
One possible fix is to use all events, but
modify the definition (\ref{eq:mct})
to approximately compensate for the transverse 
$\vec{U}_T$ boost \cite{Polesello:2009rn}.
One then recovers a distribution whose endpoint is 
still given by (\ref{MCTmaxPT0}). 
Alternatively, one could stick to the original 
$M_{CT}$ variable, and simply account for the $U_T$ 
dependence of its endpoint as
\begin{equation}
M_{CT}^{max}(U_T)=\sqrt{m_1^2+m_2^2+2m_1 m_2
\cosh\left(2\eta+\zeta_1+\zeta_2\right)}
\label{MCTmax}
\end{equation}
where $\zeta_i$ were already defined in (\ref{zetadef}), and
\begin{equation}
\sinh \eta \equiv \frac{U_T}{2M_p},
\quad
\cosh \eta \equiv \sqrt{1+\frac{U_T^2}{4M_p^2}}\, .
\end{equation}

Our approach here is to utilize the one-dimensional projections
from eqs.~(\ref{pitpar},\ref{pitperp}) and construct 
one-dimensional analogues of the $M_{CT}$ variable
\begin{eqnarray}
M_{CT_\perp} &\equiv& \sqrt{m_1^2+m_2^2+2\left(e_{1T_\perp}e_{2T_\perp}
                +\vec p_{1T_\perp}\cdot \vec p_{2T_\perp}\right)},~~~~
\label{MCTperp}
\\
M_{CT_\parallel}&\equiv&
\sqrt{m_1^2+m_2^2+2\left(e_{1T_\parallel}e_{2T_\parallel}
                +\vec p_{1T_\parallel}\cdot \vec p_{2T_\parallel}\right)},  
\label{MCTparallel}
\end{eqnarray}
where the corresponding ``transverse energies'' are 
\begin{equation}
e_{iT_\perp}     \equiv \sqrt{m_i^2+|\vec p_{iT_\perp}|^2},
\quad
e_{iT_\parallel} \equiv \sqrt{m_i^2+|\vec p_{iT_\parallel}|^2}.
\end{equation}
The benefit of the decomposition (\ref{MCTperp},\ref{MCTparallel})
is that one gets ``two for the price of one", i.e.~two independent and
complementary variables instead of the single variable (\ref{eq:mct}).

The variable $M_{CT_\perp}$ in particular
is very useful for our purposes. To illustrate the basic idea, 
it is sufficient to consider the most common case, 
where $V_i$ is approximately massless ($m_i=0$),
as the leptons in our $t\bar{t}$ example.
A crucial property of $M_{CT_\perp}$ is that its endpoint
is independent of $U_T$:
\begin{equation}
M_{CT_\perp}^{max} =
\frac{M_p^2-M_c^2}{M_p}, \quad \forall\ U_T.
\label{MCTperpmax}
\end{equation}
In fact the whole $M_{CT_\perp}$ distribution is insensitive to $U_T$:
\begin{equation}
\frac{\mathrm{d} N}{\mathrm{d}M_{CT_\perp}}
=
N_{0_\perp}\, \delta(M_{CT_\perp})
+
\left(N_{tot}-N_{0_\perp}\right)
\frac{\mathrm{d} \bar{N}}{\mathrm{d}M_{CT_\perp}},
\label{dN}
\end{equation}
where $N_{0_\perp}$ is the number of events in the zero bin
$M_{CT_\perp}=0$. Using phase space kinematics,
we find that the shape of the remaining (unit-normalized) 
zero-bin-subtracted distribution is simply given by
\begin{equation}
\frac{\mathrm{d}\bar{N}}{\mathrm{d}\hat{M}_{CT_\perp} } \equiv 
- 4\, \hat{M}_{CT_\perp}
  \ln \hat{M}_{CT_\perp} 
\label{dNbar}
\end{equation}
in terms of the unit-normalized $M_{CT_\perp}$ variable
\begin{equation}
\hat{M}_{CT_\perp} \equiv \frac{M_{CT_\perp}}{M^{max}_{CT_\perp}}.
\end{equation}

\begin{figure}[t!]
\includegraphics[width=7.0cm]{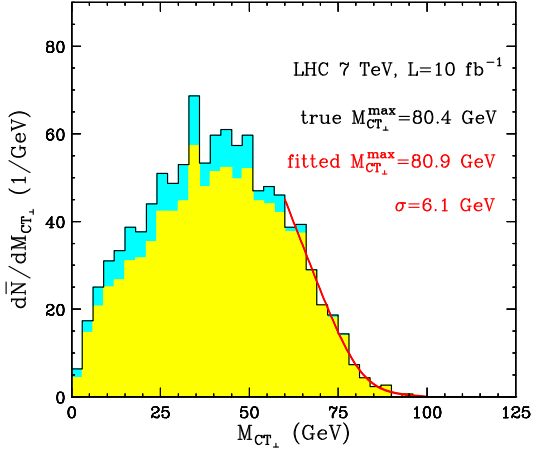}
\caption{\label{fig:perpPGS} Zero-bin subtracted 
$M_{CT_\perp}$ distribution after cuts, for
$t\bar{t}$ dilepton events. The yellow (lower) portion is our signal,
while the blue (upper) portion shows $t\bar{t}$ combinatorial background
with isolated leptons arising from $\tau$ or $b$ decays.
}
\end{figure}

The observable $M_{CT_\perp}$ distribution for 
our $t\bar{t}$ example is shown in Fig.~\ref{fig:perpPGS}, 
for $10\ {\rm fb}^{-1}$ of LHC data at 7 TeV. 
Events were generated with PYTHIA \cite{Sjostrand:2006za}
and processed with the PGS detector simulator \cite{PGS}.
We apply standard background rejection cuts as follows
\cite{Khachatryan:2010ez}: we require two isolated, opposite sign leptons 
with $p_{iT}>20$ GeV, $m_{\ell^+\ell^-}>12$ GeV,
and passing a $Z$-veto $|m_{\ell^+\ell^-}-M_Z|>15$ GeV;
at least two central jets with $p_T>30$ GeV and $|\eta|<2.4$;
and a $\met$ cut of $\met>30$ GeV ($\met>20$ GeV) for events with same 
flavor (opposite flavor) leptons. We also demand at least 
two $b$-tagged jets, assuming a flat $b$-tagging efficiency of $60\%$.
With those cuts, the SM background from other processes is
negligible \cite{Khachatryan:2010ez}.

Fig.~\ref{fig:perpPGS} demonstrates that the $M_{CT_\perp}$
endpoint can be measured quite well. Since the theoretically 
predicted shape (\ref{dNbar}) is distorted by the cuts,
we use a linear slope  with Gaussian smearing, and fit
for the endpoint and the resolution parameter.
We find $M_{CT_\perp}^{max}=80.9$ GeV
(compare to the true value $M_{CT_\perp}^{max}=80.4$ GeV),
which gives one constraint (\ref{MCTperpmax}) among $M_p$ and $M_c$.
At this point, a second, independent 
constraint can in principle be obtained from 
an analogous measurement of the $M_{CT}^{max}$ 
endpoint (\ref{MCTmax}) at a fixed value of $U_T$
(resulting in loss in statistics), after which the two masses can be found from
\begin{eqnarray}
M_p&=&\frac{U_T\, M_{CT}^{max}(U_T)\, M_{CT_\perp}^{max}}
{(M_{CT}^{max}(U_T))^2-(M_{CT_\perp}^{max})^2},
\label{Mp} \\
M_c&=&\sqrt{M_p\left(M_p-M_{CT_\perp}^{max}\right)}.
\label{Mc}
\end{eqnarray}
However, the orthogonal 
decomposition (\ref{MCTperp},\ref{MCTparallel})
offers another approach, which we pursue in the last step.

\begin{figure}[t]
\includegraphics[width=8.0cm]{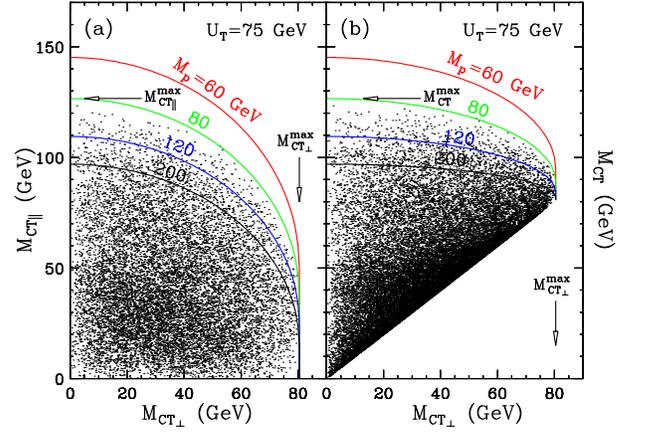}
\caption{\label{fig:parallel} 
Scatter plots of (a) $M_{CT_\perp}$ versus $M_{CT_\parallel}$
and (b) $M_{CT_\perp}$ versus $M_{CT}$, for
a fixed representative value $U_T=75$ GeV.
The solid lines show the corresponding 
boundaries defined in (\ref{MCTparhi}) 
and (\ref{MCThi}),
for the correct value of $M_{CT_\perp}^{max}$ 
and several different values of $M_p$ as shown.
}
\end{figure}

{\em Step III. Fitting to kinematic boundary lines.}
It is known that two-dimensional correlation plots reveal a lot more 
information than one-dimensional projected histograms
\cite{boundary,Matchev:2009iw}. To this end, consider the scatter plot of
$M_{CT_\perp}$ vs $M_{CT_\parallel}$ in Fig.~\ref{fig:parallel}(a), 
where for illustration we used 10,000 events at the parton level.
For a given value of $M_{CT_\perp}$,
the allowed values of $M_{CT_\parallel}$ are bounded by
\begin{equation}
M_{CT_\parallel}^{(lo)}(M_{CT_\perp})
\le M_{CT_\parallel}\le 
M_{CT_\parallel}^{(hi)}(M_{CT_\perp}),
\end{equation}
where $M_{CT_\parallel}^{(lo)}(M_{CT_\perp})= 0$ and 
\begin{equation}
M_{CT_\parallel}^{(hi)}(M_{CT_\perp}) = M^{max}_{CT_\perp}
\left( \sqrt{1-\hat{M}_{CT_\perp}^{2}}\cosh \eta + \sinh \eta\right).~~~
\label{MCTparhi}
\end{equation}
Fig.~\ref{fig:parallel}(a) reveals that the
endpoint $M_{CT_\parallel}^{max}$ 
of the one-dimensional $M_{CT_\parallel}$
distribution is obtained at
$M_{CT_\perp}=0$
\begin{eqnarray}
M_{CT_\parallel}^{max} &=&  M_{CT_\parallel}^{(hi)}(0) 
= M^{max}_{CT_\perp} (\cosh \eta + \sinh \eta) \nonumber
\\ 
&=& 
\frac{1}{2}\left(1-\frac{M_c^2}{M_p^2}\right)
\left(\sqrt{4M_p^2+U_T^2}+U_T\right) .
\label{MCTparallelmax}
\end{eqnarray}
Notice that events in the zero bins $M_{CT_\perp}=0$ and 
$M_{CT_\parallel}=0$ fall on one of the axes and cannot 
be distinguished on the plot. 

Now consider the scatter plot of 
$M_{CT_\perp}$ vs $M_{CT}$ shown in Fig.~\ref{fig:parallel}(b).
$M_{CT}$ is similarly bounded by
\begin{equation}
M_{CT}^{(lo)}(M_{CT_\perp})
\le M_{CT}\le 
M_{CT}^{(hi)}(M_{CT_\perp}),
\end{equation}
where this time $M_{CT}^{(lo)}(M_{CT_\perp}) = M_{CT_\perp}$ and
\begin{equation}
M_{CT}^{(hi)}(M_{CT_\perp}) = M^{max}_{CT_\perp}
\left(\cosh \eta + \sqrt{1-\hat{M}_{CT_\perp}^{2}} \sinh \eta\right).
\label{MCThi}
\end{equation}
We see that the endpoint $M_{CT}^{max}$ of the 
one-dimensional $M_{CT}$ distribution 
is also obtained for $M_{CT_\perp}=0$:
\begin{equation}
M_{CT}^{max} =  M_{CT}^{(hi)}(0) 
= M^{max}_{CT_\perp} (\cosh \eta + \sinh \eta)
= M_{CT_\parallel}^{max}.
\label{MCTmax2}
\end{equation}

\begin{figure}[t!]
\includegraphics[width=7.5cm]{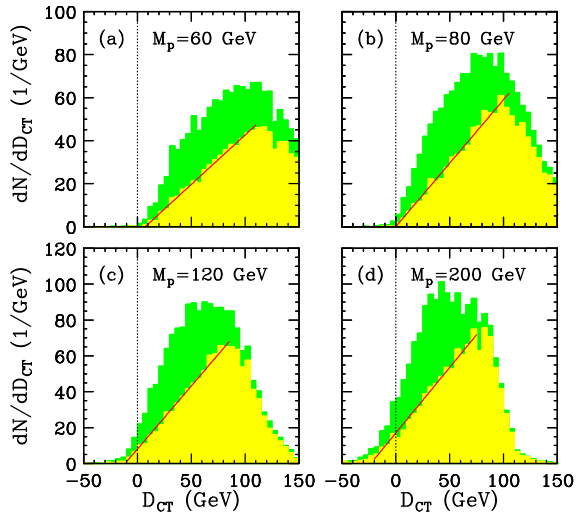}
\caption{\label{fig:dmct} $D_{CT}$ distributions 
for four different values of $M_p$ (and $M_c$ given from (\ref{Mc})).
The yellow (light shaded) histograms use
only events in the zero bin 
$M_{CT_\perp}=0$.
The red solid lines show linear
binned maximum likelihood fits.
}
\end{figure}

\begin{figure}[htb]
\includegraphics[width=7.5cm]{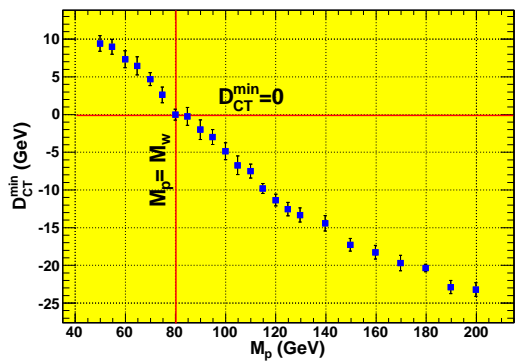}
\caption{\label{fig:dct} Fitted values of 
$D_{CT}^{min}$ as a function of $M_p$. 
}
\end{figure}

Fig.~\ref{fig:parallel} reveals a conceptual problem with one-dimensional 
projections. While {\em all} points in the vicinity of the boundary lines
(\ref{MCTparhi}) and (\ref{MCThi}) are sensitive to the masses,
the $M_{CT_\perp}^{max} $ endpoint is extracted mostly from 
events with $M_{CT_\perp}\sim M_{CT_\perp}^{max} $,
while the  $M_{CT_\parallel}^{max} $ and $M_{CT}^{max} $
endpoints are extracted mostly from the
events with $M_{CT_\perp}\sim 0 $.
The events near the boundary, but with {\em intermediate} values
of $M_{CT_\perp} $, will not enter efficiently either one of these endpoint determinations.

So how can one do better, given the knowledge of the boundary line (\ref{MCThi})?
In the spirit of \cite{Kim:2009si}, 
we define the signed distance 
to the corresponding boundary,
e.g.
\begin{equation}
D_{CT}(M_p,M_c) \equiv M_{CT}^{(hi)}(M_{CT_\perp},U_T,M_p,M_c) -M_{CT} 
\nonumber
\end{equation}
and similarly for $D_{CT_\parallel}$.
The key property of this variable is that for the correct values
of $M_p$ and $M_c$, 
its lower endpoint $D_{CT}^{min}$ 
is exactly zero (see Fig.~\ref{fig:dmct}(b)):
\begin{equation}
D_{CT}^{min}(M_p,M_c) =0.
\label{DCTmin}
\end{equation}
In that case the boundary line provides a perfectly 
snug fit to the scatter plot --- notice the green boundary line marked ``80" in 
Fig.~\ref{fig:parallel}(b).
While in general eq.~(\ref{DCTmin}) 
represents a two-dimensional fit to $M_p$ and $M_c$, in
practice one can already use the $M_{CT_\perp}^{max}$ measurement to
reduce the problem to a single degree of freedom, e.g.
the parent mass $M_p$, as presented in Figs.~\ref{fig:parallel} and
\ref{fig:dmct}. We see that the correct parent mass $M_p=80$ GeV
provides a perfect envelope, for which $D_{CT}^{min}=0$.
If, on the other hand, $M_p$ is too low, a gap develops between 
the outlying points in the scatter plot and their expected boundary,
which results in  $D_{CT}^{min}>0$.
Conversely, if $M_p$ is too high, some of the outlying points 
from the scatter plot fall outside the boundary and have $D_{CT}<0$,
leading to  $D_{CT}^{min}<0$, as seen in Fig.~\ref{fig:dmct}(c,d).
The resulting fit for $D_{CT}^{min}$ as a function of $M_p$ 
from our PGS data sample is shown in Fig.~\ref{fig:dct}, which suggests
that a $W$ mass measurement  at the level of a few percent might be viable.

{\em Acknowledgments.} 
This work is supported in part by a
US Department of Energy grant DE-FG02-97ER41029.


\begin{thebibliography}{99}

\bibitem{imass}
  I.~Hinchliffe {\em et al.}, 
  Phys.\ Rev.\  D {\bf 55}, 5520 (1997);
  B.~C.~Allanach {\em et al.}, 
  JHEP {\bf 0009}, 004 (2000);
  B.~K.~Gjelsten, D.~J.~Miller and P.~Osland,
  JHEP {\bf 0412}, 003 (2004).

\bibitem{Matchev:2009iw}
  K.~T.~Matchev, F.~Moortgat, L.~Pape and M.~Park,
  JHEP {\bf 0908}, 104 (2009).
  
\bibitem{exactreco}
  K.~Kawagoe, M.~M.~Nojiri and G.~Polesello,
  Phys.\ Rev.\  D {\bf 71}, 035008 (2005);
  H.~C.~Cheng {\em et al.}, 
  Phys.\ Rev.\ Lett.\  {\bf 100}, 252001 (2008).

\bibitem{Khachatryan:2010ez}
  V.~Khachatryan {\it et al.}  [CMS Collaboration],
  arXiv:1010.5994 [hep-ex].
  
\bibitem{approxreco}
  C.~G.~Lester and D.~J.~Summers,
  Phys.\ Lett.\  B {\bf 463}, 99 (1999);
  A.~Barr, C.~Lester and P.~Stephens,
  J.\ Phys.\ G {\bf 29}, 2343 (2003).

\bibitem{kink}
  A.~J.~Barr, B.~Gripaios and C.~G.~Lester,
  JHEP {\bf 0802}, 014 (2008);
  M.~Burns, K.~Kong, K.~T.~Matchev and M.~Park,
  JHEP {\bf 0903}, 143 (2009).

\bibitem{Matchev:2009fh}
  K.~T.~Matchev, F.~Moortgat, L.~Pape and M.~Park,
  Phys.\ Rev.\  D {\bf 82}, 077701 (2010);
  P.~Konar, K.~Kong, K.~T.~Matchev and M.~Park,
  Phys.\ Rev.\ Lett.\  {\bf 105}, 051802 (2010).

\bibitem{Tovey:2008ui}
  D.~R.~Tovey,
  JHEP {\bf 0804}, 034 (2008).

\bibitem{Polesello:2009rn}
  G.~Polesello and D.~R.~Tovey,
  JHEP {\bf 1003}, 030 (2010).

\bibitem{Sjostrand:2006za}
  T.~Sjostrand, S.~Mrenna and P.~Skands,
  JHEP {\bf 0605}, 026 (2006).

\bibitem{PGS}
http://www.physics.ucdavis.edu/$\sim$conway/research/\\
software/pgs/pgs4-general.htm

\bibitem{boundary}
  D.~Costanzo and D.~R.~Tovey,
  JHEP {\bf 0904}, 084 (2009);
  M.~Burns, K.~T.~Matchev and M.~Park,
  JHEP {\bf 0905}, 094 (2009).

\bibitem{Kim:2009si}
  I.~W.~Kim,
  Phys.\ Rev.\ Lett.\  {\bf 104}, 081601 (2010).

\end{thebibliography}
\end{document}